\documentclass[11pt,a4paper]{article}
\pdfoutput=1 
\usepackage{jheppub}

\usepackage{amsmath,amssymb}
\usepackage{graphicx}
\usepackage{subcaption}  
\usepackage{xcolor}     
\usepackage{hyperref}   
\usepackage{slashed}
\usepackage{cool}
\allowdisplaybreaks

\providecommand{\sumintB}{\ensuremath{\int \mkern-24mu \sum_{k_\parallel}}}

\providecommand{\dkp}{\ensuremath{\int\,\frac{d^2k_\bot}{\left(2\pi\right)^2}}}

\providecommand{\sumrs}{\ensuremath{\sum_{rs}}}
\providecommand{\qfb}{\ensuremath{\left\lvert q_f \mathcal{B}\right\rvert}} 

\title{Meson spectral function and screening masses in magnetized
  quark gluon plasma}

\author[]{Purnendu Chakraborty}
\affiliation[]{Department of Physics, Basirhat College, Basirhat, WB, India.}
\emailAdd{purnendu.chakraborty@gmail.com} 

\keywords{Heavy Ion Collision,  Quark Gluon Plasma, Magnetic field, Spectral function}

\abstract{
We calculate the spectral function in pseudoscalar (scalar) channel 
in the high temperature phase of QCD in presence of a 
background magnetic field. Spatial and temporal screening masses are determined from the long
distance behavior of the corresponding correlation functions.        
}    

\begin{document}
\maketitle 
\flushbottom

Under extreme conditions of temperature and  baryon density, strongly
interacting matter of quantum chromodynamics (QCD) liberates a large
number of degrees of freedom indicating a phase transition to a deconfined,
quasi-ideal state known as quark gluon plasma (QGP)~\cite{Aoki:2006we}.  
There is an overall consensus that heavy ion collision experiments have
shown us glimpses of such a state of matter. It is known for sometime that
ultra-relativistic motion of charged particles create an
intense magnetic field in the early stage of non-central heavy ion collision. 
The energy scale of the magnetic field thus generated is comparable
 with the characteristic scale of QCD, for example,    
$\mathcal{B} \sim m_\pi^2/e$ at RHIC and could be as high as  
$\mathcal{B} \sim 10 m_\pi^2/e$ at LHC~\cite{Kharzeev:2007jp,Skokov:2009qp,
Bzdak:2011yy, Voronyuk:2011jd, Deng:2012pc}. Here $m_\pi$  is pion mass in vacuum
 and $e$  is  charge of proton. 
An external magnetic field modifies the QCD vacuum and entails a 
rich spectrum of phenomena - chiral magnetic effect 
(CME)~\cite{Kharzeev:2004ey,Kharzeev:2007jp,Buividovich:2009wi,
Fukushima:2010vw}, chiral vortical effect (CVE), magnetic catalysis, modification of the phase
diagram~\cite{Bali:2011qj,Bali:2012zg,Fraga:2012fs} and so on. 
There are tremendous amount of activities, both theoretical and experimental, going on to
understand the properties of QCD matter under an external magnetic
field~\cite{Kharzeev:2013jha}. Apart from QCD, effects induced by an external
magnetic field is important in astrophysics~\cite{Harding:2006qn}, cosmology~\cite{Vachaspati:1991nm},
physics beyond standard model~\cite{Gies:2006ca}, or condensed matter physics~\cite{Miransky:2015ava}.

Hadronic correlation functions are useful objects to understand the
intricate dynamics of QCD~\cite{Shuryak:1993kg,DeTar:1987ar,DeTar:1987xb}.
Spectral densities of correlation functions encode information of  
in-medium hadron properties, transport coefficients and electromagnetic
emissivity from the hot and dense plasma.   
Mesonic spectral functions at finite temperature have been calculated 
in the literature using analytic methods~\cite{Florkowski:1993bq, Karsch:2000gi,
  Alberico:2006wc, Vepsalainen:2007ke, Burnier:2012ze} or numerical 
  simulations of lattice QCD~\cite{Karsch:2003wy,
  Aarts:2005hg,Brandt:2012jc,Cheng:2010fe}. Hadronic
  correlators in 
a background magnetic field have also been studied  in different settings,  see  
\cite{Sadooghi:2016jyf, Avancini:2016fgq, 
Buividovich:2010qe,Buividovich:2010tn,Ghosh:2017rjo, Bandyopadhyay:2017raf,
Bandyopadhyay:2016fyd, 
Ghosh:2016evc,Mukherjee:2017dls} 
for latest development in the field.     

The purport of the present paper is to discuss the modification of  mesonic 
spectral densities in the high temperature deconfined phase of QCD. We shall 
work to $\mathcal{O}\left(\alpha_s^0\right)$ in the strong coupling constant 
albeit the effect of magnetic field, by construction, is included to all
orders. Neglect of QCD radiative corrections provides  a clean benchmark to
understand the effect of a magnetic field on the the propagation of mesons and
 it serves to define an appropriate  starting point for a refined 
analysis with higher order QCD effects systematically embedded. 
On the  phenomenology side, such an approximation  may be quite relevant 
at the top LHC energy and Future circular collider (FCC)
respectively. 

For brevity, we shall consider only neutral pseudo-scalar (scalar) mesons of 
chiral quarks in this paper.  We also assume that  mesons are
composed only one kind of quark flavor which will be either $u$ or
$d$.
Thus our mesons are not physical mesons but these states can be 
constructed in the laboratory of lattice QCD~\cite{Bali:2017ian}.

Analytic studies in a  background magnetic field have rarely been pushed beyond 
one loop and even at one loop order the calculations are arranged for some special
configurations of fields most of the time. If the magnetic
field dominates other scales in the problem, then it makes sense
to place the charged particles in the lowest Landau level
(LLL) because states at higher Landau levels are too heavy to be excited.  The
beauty of the LLL approximation is that it allows complete separation of
motion along the direction parallel to the magnetic field and 
gyromagnetic motion in the transverse space. Furthermore, it allows much
simpler tensorial structure of $n$ point functions and easy Gaussian 
integrations over transverse momenta of the virtual  particles. In
general, long distance properties are
sensitive to the LLL. However, barring special observable like chiral magnetic 
current or spin polarization where only LLL contribute, restriction to 
LLL brings in uncertainty in the calculation. Effect of higher
landau levels are   accommodated in the loop calculation either by choosing  
special direction of propagation with respect to the external magnetic field 
$\left(\vec{p} \parallel \, \textrm{or}\,  \bot\,  \vec{\mathcal{B}}\right)$ 
or through a partial resummation of arbitrary Landau
levels~\cite{Kuznetsov:680986} in the strong field limit. 
Another kind of resummation of Landau levels is applicable when the magnetic
field is weaker than  pertinent mass scales in the
problem. Here the resummation is equivalent to the expansion of the propagator
in powers of the magnetic field~\cite{Chyi:1999fc} which is mostly  
useful to calculate the high frequency tail of the 
massive correlators~\cite{Machado:2013yaa,Cho:2014loa}.~\footnote{Such
expansion coincides with the 
operator product expansion which  has been widely 
used in the context of QCD sum rule calculations in nonperturbative
background of color electromagnetic fields. For massless particle, the
exercise of OPE needs certain care. The point is that the propagator becomes
increasingly sensitive to the infrared as one moves to higher order in the
expansion which is translated in the infrared sensitivity of the
correlator. To save the whole scheme from doom, one needs to absorb 
long distance divergences in the definition of condensates leaving short
distance contributions in coefficient functions. The procedure
is known for QCD~\cite{Broadhurst:1984rr,Zschocke:2011aa, Grozin:1994hd}
  and we have explicitly
checked in the case of electromagnetic correlator that it works in presence of a
background $\textrm{U}(1)$ field too. We do not discuss this type of
calculation here which in a sense is redundant when the complete result is
known. The interested reader may see~\cite{Patkos:1979zg} for a case in this
point.}

The plan of the present paper is as follows. Notations used in the paper and formalism are
introduced in Sec.\ref{formalism}. We derive an analytic expression of
mesonic spectral density for entire 
momentum range in
Sec.\ref{spectral_function}. This is the central result 
of the paper. We discuss asymptotic limits of different correlators and find
corresponding screening masses in Sec.\ref{screening_mass}.

\section{Formalism} \label{formalism} 
For definiteness, we assume a spatio-temporally constant magnetic field
along the $z$ direction. 
 The hadronic current is given by $J_h \left(\tau, \vec{x}\right) =
\bar{q}\left(\tau, \vec{x}\right) \Gamma_h q\left(\tau, \vec{x}\right)$. Here, 
$\Gamma_h = 1, \gamma_5 $ for 
scalar (S), pseudo-scalar (PS) respectively. 

\subsection*{Correlation function} 
Since the magnetic field breaks
the isotropy of space, the in-medium correlation functions depend 
on longitudinal ($z$) and transverse ($x_\bot$) coordinates separately, 
\begin{eqnarray}
\chi_h \left(\tau, x_\bot, z\right) &=& \left\langle 
J_h\left(\tau, x_\bot, z\right) 
J_h^\dagger\left(0, 0_\bot, 0\right) \right\rangle_\beta\,, \nonumber \\
&=&\frac{1}{\beta} \sum_{n = -\infty}^{+\infty} \int
\frac{d^3p}{\left(2\pi\right)^3} e^{-i\left(\omega_n \tau - \vec{p}\cdot\vec{x}
\right)}\chi_h\left(\omega_n, p_\bot, p_z\right)\,. 
\label{corr_fn_def1}
\end{eqnarray}
The spectral density $\sigma_h \left(\omega, p_\bot, p_z\right)$, upto
possible subtractions, is defined as  
\begin{eqnarray} 
\chi_h\left(\omega_n, p_\bot, p_z\right) &=& \int_{-\infty}^{+\infty} du 
\frac{\sigma_h \left(u, p_\bot, p_z\right)}{u - i \omega_n}\,, \nonumber \\
\Rightarrow \sigma_h \left(\omega, p_\bot, p_z\right) &=& \operatorname{\Im}
\chi_h\left(i\omega_n = \omega + i \epsilon, p_\bot, p_z\right)\,.
\end{eqnarray}  
Correlation functions of interest can be expressed in
terms of the spectral function.

\begin{itemize} 
\item Temporal meson correlation function : 
\begin{eqnarray} 
\label{temporal_correlator} 
\chi_h^\tau\left(\tau, p_\bot, p_z\right) &=& \int d^3x\,
e^{-i\vec{p}\cdot\vec{x}} \chi_h\left(\tau, x_\bot, z\right)\,,  \nonumber \\
&=& \frac{1}{\beta} \sum_{n = -\infty}^{+\infty} e^{-i\omega_n \tau
}\chi_h \left(\omega_n, p_\bot, p_z\right)\,,  \nonumber \\ 
&=& \int_0^\infty d\omega \, \sigma_h \left(\omega, p_\bot, p_z\right) 
K \left(\omega,\tau\right)\,,
\end{eqnarray}  
where, $K\left(\omega, \tau\right) = \frac{\cosh{\omega\left(\tau
-\frac{\beta}{2}\right)}}{\sinh{\frac{\beta\omega}{2}}}$.

\item Longitudinal  correlation function : 
\begin{eqnarray} 
\label{longitidinal_correlator}
\chi_h^z\left(z\right) &=& \int_0^\beta d\tau \int d^2x_\bot \,
\chi_h\left(\tau, x_\bot, z\right)\,,\nonumber \\
&=& \int_{-\infty}^{+\infty} \frac{dp_z}{2\pi} e^{ip_z z} \int_{0}^{\infty}
d\omega \frac{\sigma_h
  \left(\omega, p_\bot = 0, p_z\right)}{\omega}\,. 
\end{eqnarray} 
 
\item  Transverse plane correlation function
\begin{eqnarray} 
\label{transverse_correlator}
\chi_h^{xy}\left(\vec{x}_\bot\right)  &=& \int_0^\beta d\tau \int dz \, 
\chi_h\left(\tau, x_\bot, z\right) \,,\nonumber \\
&=& \int \frac{d^2p_\bot}{\left(2\pi\right)^2} e^{i\vec{p}_\bot\cdot \vec{x}_\bot} \int_{0}^{\infty} d\omega \frac{\sigma_h
  \left(\omega, p_\bot, p_z = 0 \right)}{\omega} \,.
\end{eqnarray}  
\end{itemize}

If the spectrum of the theory is characterized by simple poles, 
then the corresponding Fourier transforms will feature exponential fall off 
at long distance. The inverse of the characteristic range of the 
correlation function is called the screening mass. In the
chiral limit, temporal and spatial screening masses 
are equal in free theory and  given by $m_{\rm scr} = 2 \pi T$ which has 
simple interpretation as arising due to independent
propagation of two quarks. If we neglect QCD effects, equality of the
screening masses still hold for longitudinal and 
temporal directions even when a magnetic field is present. In the lowest order 
of perturbation theory, the screening masses are in fact independent 
of magnetic field and given by free theory value. This is a consequence of the
fact that 
long distance properties of the correlator are determined by LLL which is 
independent of $\mathcal{B}$.
It will be shown
later that the transverse plane 
correlator shows a Gaussian fall off at large distance with a  mass
scale $\sim \sqrt{\qfb}$, where $q_f$ is the charge of the flavor $f$.
This is not a screening behavior per se, but
reminiscent of the magnetic confinement in the transverse plane with a
characteristic scale $ r_\perp \sim 1/\sqrt{\qfb}$.       

\subsection*{Fermion Propagator} 
The exact charged fermion propagator in a homogeneous external field 
can be written as
\begin{equation}
\label{schwinger_prop1}
\widetilde{S_f}\left(x, x^\prime\right) = e^{\lambda \left(x, x^\prime\right)} 
\int \frac{d^4k}{\left(2\pi\right)^4} \, e^{-i k \left(x - x^\prime\right)}
S_f \left(k\right)\,.
\end{equation} 
Here $S_f\left(k\right)$ is the translation and gauge invariant part of 
the fermion propagator in a background  potential
$A_{\mu}^{\rm ext}\left(x\right)$. 
The holonomy factor $\lambda\left(x, x^\prime\right)$ breaks gauge and 
translation invariance. Explicit form of $\lambda$ is not important here, 
it drops out in a 
gauge invariant calculation.\footnote{For constant 
  electromagnetic fields, holonomy factors from the propagators in the two 
vertex fermion loop cancel in the correlation function for neutral
mesons. This is not the case  for charged mesons.} $S_f$ can be 
decomposed as sum over the discrete Landau 
levels~\cite{Chodos:1990vv, Gusynin:1995nb},
\begin{subequations}
\label{prop_chodos}
\begin{align}
i S_f\left(k\right) &= i e^{- \rho}
\sum_{n = 0}^\infty \left(-1\right)^n D_n \left(k_\parallel,
  k_\perp\right)
\Delta_f\left(k_\parallel, \epsilon_n\right)\,, \label{ce1}\\
D_n \left(k_\parallel, k_\perp\right) &= 2\left(\slashed{k}_\parallel + m
\right) \left(\mathcal{P}^{-} 
L_n\left(2 \rho\right) - \mathcal{P}^{+} 
L_{n-1}\left(2 \rho\right)\right) - 4 \slashed{k}_\perp L_{n-1}^1  \left(2 \rho\right)\,, \label{ce2}\\
\Delta_f\left(k_\parallel, \epsilon_n\right) &= \left(k_\parallel^2 - 
\left(\epsilon_n\right)^2\right)^{-1} = \left(k_\parallel^2 - m_f^2 - 2 n
\left|q_f \mathcal{B}\right|\right)^{-1}\,.\label{ce3}
\end{align}
\end{subequations}
Our notation here is as follows : the four vectors are decomposed into 
components parallel and perpendicular to magnetic field,  $a^\mu =
a^\mu_\parallel + a^\mu_\perp$, where $a^\mu_\parallel = \left(a^0, 0, 0, a^3\right)$ and 
$a^\mu_\perp = \left(0, a^1, a^2, 0\right)$. The metric tensor is written as 
as $g^{\mu\nu} = g^{\mu\nu}_\parallel + g^{\mu\nu}_\perp$, where
$g^{\mu\nu}_\parallel = \mathrm{diag}\left(1, 0, 0, -1\right)$ and 
$g^{\mu\nu}_\parallel = \mathrm{diag}\left(0, -1, -1, 0\right)$. The scalar
product naturally splits as   $a \cdot b = (a \cdot b)_\parallel + (a \cdot b)_\perp$ 
where $(a \cdot b)_\parallel= a^0b^0 - a^3 b^3$ and $(a \cdot b)_\perp 
= -\left(a^1b^1 + a^2 b^2\right)$. Let us also note that $q_f$ and $m_f$ 
are charge and mass of the fermion respectively.  
We have taken $\rho = \mathbf{k}_\perp^2/{\left|q_f \mathcal{B}\right|}$.
$P^\pm = \frac{1}{2}\left(1 \pm i \gamma^1
\gamma^2 \mathrm{sgn}\left(\left|q_f \mathcal{B}\right|\right)\right)$ are 
spin projection operators along the magnetic field direction.  $L_n^\alpha$ are 
associated Laguerre polynomials. By definition, $L_n = 0$ if $n < 0$.

\section{Spectral Function} \label{spectral_function}

\begin{figure}[!htbp]
\centering{
\includegraphics[width=3cm]{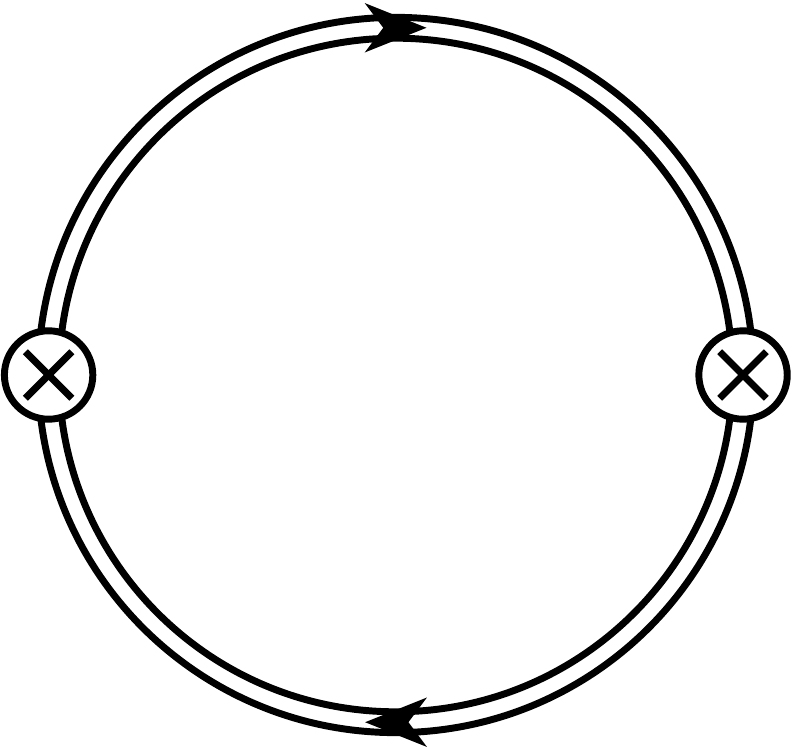}
}
\caption{Magnetically confined but otherwise free meson correlator. 
Double lines represent exact fermion propagator in a magnetic field.} 
\end{figure}  
The correlation function is given by the convolution of fermion
propagators. Using the propagator \eqref{prop_chodos} pseudoscalar correlator 
can be recast in the form,   
\begin{eqnarray}
\label{chi_ps1} 
\chi_{ps} \left(\omega, p_\bot, p_z\right) &=&  - 4 N_c  \sumrs \sumintB \dkp\,  \left(-1\right)^{r+s}
e^{-\left(\rho_k + \rho_q\right)} \Delta_f\left(k_\parallel,\epsilon_r\right) 
\Delta_f\left(q_\parallel,\epsilon_s\right) \nonumber \\ 
&\times& \left[\mathcal{F}_{r,s} \left\{L_r\left(2 \rho_k\right) L_s\left(2
  \rho_q\right) + L_{r-1}\left(2 \rho_k\right) L_{s-1}\left(2
  \rho_q\right)\right\} \right. \nonumber \\ 
&& \left. - 16 \left(k \cdot q\right)_\bot 
L_{r-1}^1\left(2 \rho_k\right) L_{s-1}^1\left(2 \rho_q\right)\right] +
\mathcal{C} \nonumber \\
\text{with} \, \quad \mathcal{F}_{r,s} &=&  \left(s_\parallel - 
2 \qfb r - 2 \qfb s \right) 
\label{corr_ps}
\end{eqnarray} 
$\mathcal{C}$ represents terms without discontinuities. Here $s_\parallel =
\omega^2 - p_z^2$ and the sum integral stands for, 
\[\int \frac{d^2k_\parallel}{\left(2\pi\right)^2} \rightarrow i T \sum_{k_0} 
\int \frac{dk_z}{2\pi} = i \int \mkern-24mu
\sum_{k_\parallel}\,.
\]
The frequency sum is most conveniently done using the mixed representation of the propagator, 
\begin{equation}
\Delta_f \left(\tau\right) = T \sum_{k_0} e^{-k_0 \tau} \Delta_f
\left(k_\parallel, \epsilon_r \right)\,,  
\end{equation} 
where 
\begin{equation}
\Delta_f \left(\tau\right) = -\frac{1}{2 \omega_{r}\left(k\right)} \left[\left(1 - 
n_f \left(\omega_{r}\left(k\right)\right)e^{- \omega_{r}\left(k\right)
  \tau}\right) - n_f \left(\omega_{r}\left(k\right)\right)
e^{\omega_{r}\left(k\right) \tau} \right]\,.
\end{equation} 
Here $\omega_{r}\left(k\right) = \sqrt{k_z^2 + \epsilon_r^2}$ and $n_f$ is
Fermi-Dirac distribution function. 
Let us define 
\begin{subequations}
\begin{align} 
\mathcal{T}^{\alpha,\beta,\gamma}_{r,s} &= \left(-1\right)^{r+s} \dkp 
e^{-\left(\rho_k + \rho_q\right)} L_r^\alpha \left(2 \rho_k\right) L_s^\beta\left(2
  \rho_q\right) \left(\vec{k}_\bot\cdot\vec{q}_\bot\right)^\gamma\,, \\
\mathcal{I}_{r,s} &= - \sumintB \Delta_f\left(k_\parallel,\epsilon_r\right) 
\Delta_f\left(q_\parallel,\epsilon_s\right)\,.
\end{align}  
\end{subequations}
Then the spectral function can be written as, 
\begin{equation}
\label{spec_ps_gen} 
\sigma_{ps} \left(\omega, \vec{p}_\bot, p_z\right) = - 4 N_c  \sumrs 
\left[\mathcal{F}_{rs} \left(\mathcal{T}^{0,0,0}_{r,s} + 
\mathcal{T}^{0,0,0}_{r-1,s-1}\right) + 16 \mathcal{T}_{r-1,s-1}^{111}\right]
\Im \left(\frac{\mathcal{I}_{r,s}}{\pi} \right)\,.
\end{equation} 
Similarly the spectral function in the scalar channel can be written as, 
\begin{equation}
\label{spec_s_gen}
\sigma_{s} \left(\omega, \vec{p}_\bot, p_z\right) = - 4 N_c \sumrs 
\left[\mathcal{G}_{r,s} \left(\mathcal{T}^{0,0,0}_{r,s} + 
\mathcal{T}^{0,0,0}_{r-1,s-1}\right) + 16 \mathcal{T}_{r-1,s-1}^{111}\right]
\Im \left(\frac{\mathcal{I}_{r,s}}{\pi} \right)\,,
\end{equation} 
where, $\mathcal{G}_{rs} = \left(s_\parallel - 4 m^2 - 2 \qfb r -
2 \qfb s\right)$.
 
Expressions for $\Im \left(\mathcal{I}_{r,s}\right)$ and $\mathcal{T}$ 
integrals are worked out in appendices. Let us note that $\mathcal{I}_{r,s}$
are property of the longitudinal $\left(\parallel\right)$ space  whereas
$\mathcal{T}$ integrals belong to
transverse $\left(\perp\right)$ space. This near complete factorization 
makes the interpretation of the spectral function clear. The
correlator in \eqref{chi_ps1} can be thought of as superposition of mesonic
states with quark-antiquark pair in $(r,s)$ Landau
levels. Each such mesonic state has its own spectral density
$\sigma_{r,s}$. $\sigma_{r,s}$ consist of annihilation contribution and 
scattering contribution which is typical of a thermal medium. What is
different in the magnetized plasma is dimensional reduction. Since
discontinuity of the correlator is determined by $I_{r,s}$, the structure of
the spectral function is essentially
that of two dimensional field theory in $\left(0,3\right)$ plane. 
The gyromagnetic motion in the transverse plane does not lead to any new cut
in the energy plane.  The background
magnetic field acts just like a medium and it shifts the location of 
cuts in the energy plane by endowing  quarks an effective  mass  
$\sim \sqrt{q_f B}$. It is the quantized momentum of the charged particles in the
transverse plane which acts like a mass term for motion in the longitudinal
direction.  
 
$\mathcal{F}_{r,s}$ and $\mathcal{T}$ integrals can be explained in the same
way by comparing with the corresponding expressions in the free theory~\cite{Weldon:1983jn}. 
The magnetic field  modifies the scattering amplitudes and these
modifications are contained in the $\mathcal{T}$ integrals. $\mathcal{F}_{r,s}$
has the same interpretation as $\mathcal{I}_{r,s}$. We can simply obtain it
from free theory by dropping all reference to transverse dynamics
$\left(s \to s_{\parallel}\right)$ and augmenting the bare mass by quantized
transverse momentum $\left(m^2 \to m^2 + 2 l\qfb\right)$.

We note that spectral function of
pseudoscalar and scalar channels are degenerate in the chiral limit 
$\sigma_{ps} = \sigma_{s}$ although  fermions of
the theory became ``massive''.  Thus magnetically generated mass does not lead 
to chiral symmetry breaking, at least it is not captured in the lowest order
of perturbation theory.

The structure of the spectral function is  much simplified in two special
circumstances - 1) when momentum of the meson is aligned with the magnetic field
or 2) when the quark-antiquark pair occupy LLL. Let us pause for while to
discuss these two special cases before we disseminate the results.   

\subsection*{Spectral function for $\vec{p}_\bot = 0$.} 

Let us set $\vec{p}_\bot = 0$ in \eqref{spec_ps_gen}. 
$\mathcal{T}$ integrals now reduce to normalization integrals (see~\eqref{normalization_integral}), 
\begin{equation}
\label{lag_id}
\mathcal{T}_{r,s}^{0,0,0} = \frac{\qfb}{8\pi} \delta_{r,s}\,,\,\, 
\mathcal{T}_{r,s}^{1,1,1} = \frac{\qfb^2}{16\pi} \delta_{r,s}\,. 
\end{equation}
Substituting \eqref{lag_id} in \eqref{corr_ps}, we get 
\begin{equation}
\chi^{ps} \left(\omega, p_z\right) = -N_c \frac{\qfb}{2\pi} \sum_r \left(2 -
\delta_{r,0}\right) \mathcal{F}_{rr} \mathcal{I}_{r,r} - N_c \frac{4
  \qfb^2}{\pi} \sum_r r \mathcal{I}_{r,r}\,. 
\end{equation} 
The magnetic field dependent contribution in the first term from 
$\mathcal{F}_{rr}$ cancels similar contribution in the second term.  
The spectral function now follows as, 
\begin{equation}
\chi^{ps} \left(\omega, p_z\right) = -N_c \frac{\qfb}{2 \pi} s_\parallel
\sum_{r} \left(2 - \delta_{r,0}\right) \mathcal{I}_{r,r}\,.
\end{equation} 
Using~\eqref{eq_discont}, the spectral function follows as
\begin{eqnarray} 
\label{spec_ps_expanded} 
\sigma_{ps} \left(\omega, p_z\right) &=& N_c\frac{\qfb}{8\pi^2} \sum_r \left(2 -
\delta_{r,0}\right) \theta\left(s_\parallel - 4 \epsilon_r^2\right) \frac{1}{\sqrt{1 - \frac{4
      \epsilon_r^2}{s_\parallel}}} \left(1 - n_f \left(\omega_r^+\right) -
n_f\left(\omega_r^-\right)\right) \nonumber \\ 
&+& N_c\frac{\qfb}{8\pi^2} \sum_r \left(2 -
\delta_{r,0}\right) \theta\left(- s_\parallel\right) \frac{1}{\sqrt{1 - \frac{4
      \epsilon_r^2}{s_\parallel}}} \left(n_f \left(\tilde{\omega}_r^+\right) -
n_f\left(- \tilde{\omega}_r^-\right)\right) \nonumber \\ 
\text{where} \quad \omega_r^\pm &=& \frac{\omega}{2} \pm \frac{p_z}{2} \sqrt{1 -
  \frac{4 \epsilon_r^2}{s_\parallel}}\,,\,\,\, \tilde{\omega}_r^\pm = 
\frac{\omega}{2} \pm \frac{\lvert p_z \rvert}{2} \sqrt{1 -
  \frac{4 \epsilon_r^2}{s_\parallel}}.
\end{eqnarray} 
In the limit of very weak field, the difference in energy between adjacent
Landau levels becomes very small and $r$ in this case can be taken as
continuous variable. Replacing summation over $r$ by an integration and using
the following identities, 
\begin{eqnarray}
n_f \left(x\right) &=& \sum_{r = 1}^{\infty} \left(-1\right)^{r + 1} e^{-rx}
\,,\\
\log{\left(1 + e^{-x}\right)} &=& \sum_{r = 1}^\infty \frac{\left(-1\right)^{r
    + 1}}{r}  e^{-rx}\,,
\end{eqnarray}  
it is not difficult to show that in the limit of very weak field,  
the spectral function is approximated by free field value
\begin{equation}
\sigma_{ps} \left(\omega, p_z\right) \overset{\qfb \to 0}{\simeq}  \frac{N_c}{8\pi^2}
\frac{T}{\lvert p_z \rvert} \left[
\theta\left(s_\parallel - 4 m^2\right) 
\log{\frac{\cosh{\frac{\beta\tilde{\omega}_+}{2}}}{\cosh{\frac{\beta\tilde{\omega}_-}{2}}}}
+ \theta\left(-s_\parallel\right) \log{\frac{
\left(1 + e^{-\beta \tilde{\omega}_+}\right)}{\left(1 + e^{\beta \tilde{\omega}_-}\right)}}
\right]\,. 
\end{equation} 

\subsection*{Spectral Function in strong field limit} 
Suppose all mass scales in the problem are smaller than $\qfb$. We may
assume that quarks are occupying the lowest Landau levels, which is the
lowest energy state. The fermion propagator in LLL is obtained by setting $n =
0$ in \eqref{prop_chodos},     
\begin{equation}
\label{prop_lll} 
i S_f^{\rm LLL} \left(k\right) = 2 i e^{-\frac{\mathbf{k}_\perp^2}{\left|q_f B\right|}}
\left(\slashed{k}_\parallel + m_f \right) \mathcal{P}^{-} \Delta_f\left(k_\parallel\right)\,.
\end{equation} 
Apart from spin projection operator, $\left(\slashed{k}_\parallel + m_f \right)
\Delta_f\left(k_\parallel\right)$ is just the free particle propagator in $\parallel$
space.  We notice that in the lowest landau level, the dynamics in  $\parallel$
and $\perp$ space have been completely separated  at the propagator level. The
exponential factor in $S_f$ is interesting. After Fourier transform, it gives
a factor $\exp\left({-\frac{1}{2} \qfb x_\bot^2}\right)$ which (unlike in free
theory) is a non diverging function with maximum at $x_\bot = 0$.  
In a magnetic field, the charged particles funnel along the field lines. A
quark in LLL, gyrates in an orbit with Larmor radius $r_\bot \sim
1/\sqrt{\qfb}$. Thus starting at $x_\bot = 0$ it will land up within
region $\delta x_\bot^2 \sim 1/\qfb$. The spectral function can be obtained
from \eqref{eq_discont} and \eqref{chi0int} as, 
\begin{equation}
\label{spec_lll}
\lvert \sigma_{s} \rvert = \lvert \sigma_{ps} \rvert = \frac{\lvert q_f B\rvert}{4 \pi^2}
\frac{e^{- \frac{p_\bot^2}{2 \lvert q_f B\rvert}} }{ 
\left(\sqrt{1 - \frac{4 m^2}{s_\parallel}} \right)} \left(1 -
n_F\left(\omega_+\right) - n_f\left(\omega_-\right) \right) \theta\left(s_\parallel - 4 m_f^2\right)\,.
\end{equation}

\subsection*{Results for Spectral Function} 
\begin{figure}[htbp!]
  \begin{subfigure}[b]{0.48\textwidth}
    \includegraphics[width=\textwidth]{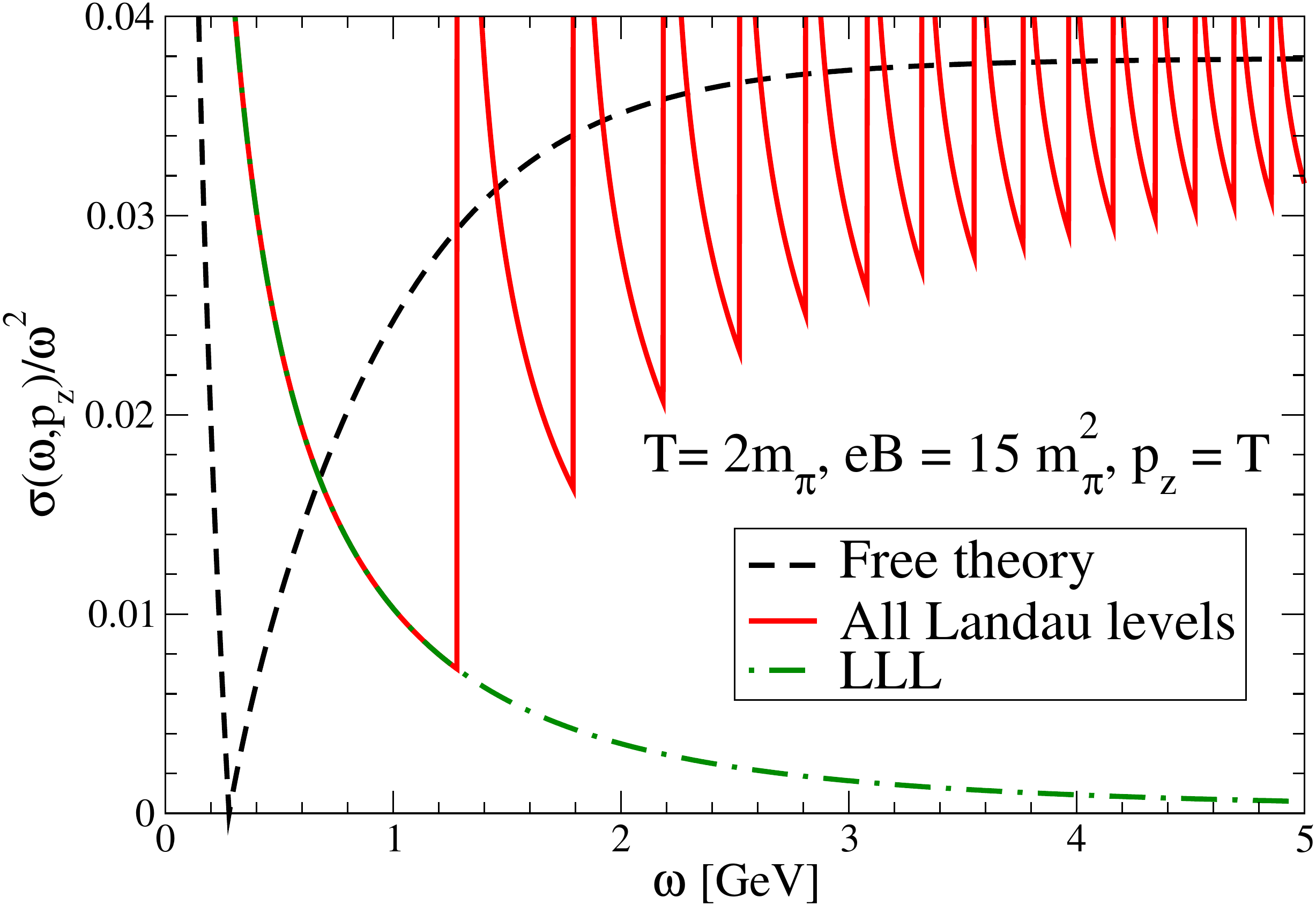}
  \end{subfigure}
  \hfill 
  \begin{subfigure}[b]{0.48\textwidth}
    \includegraphics[width=\textwidth]{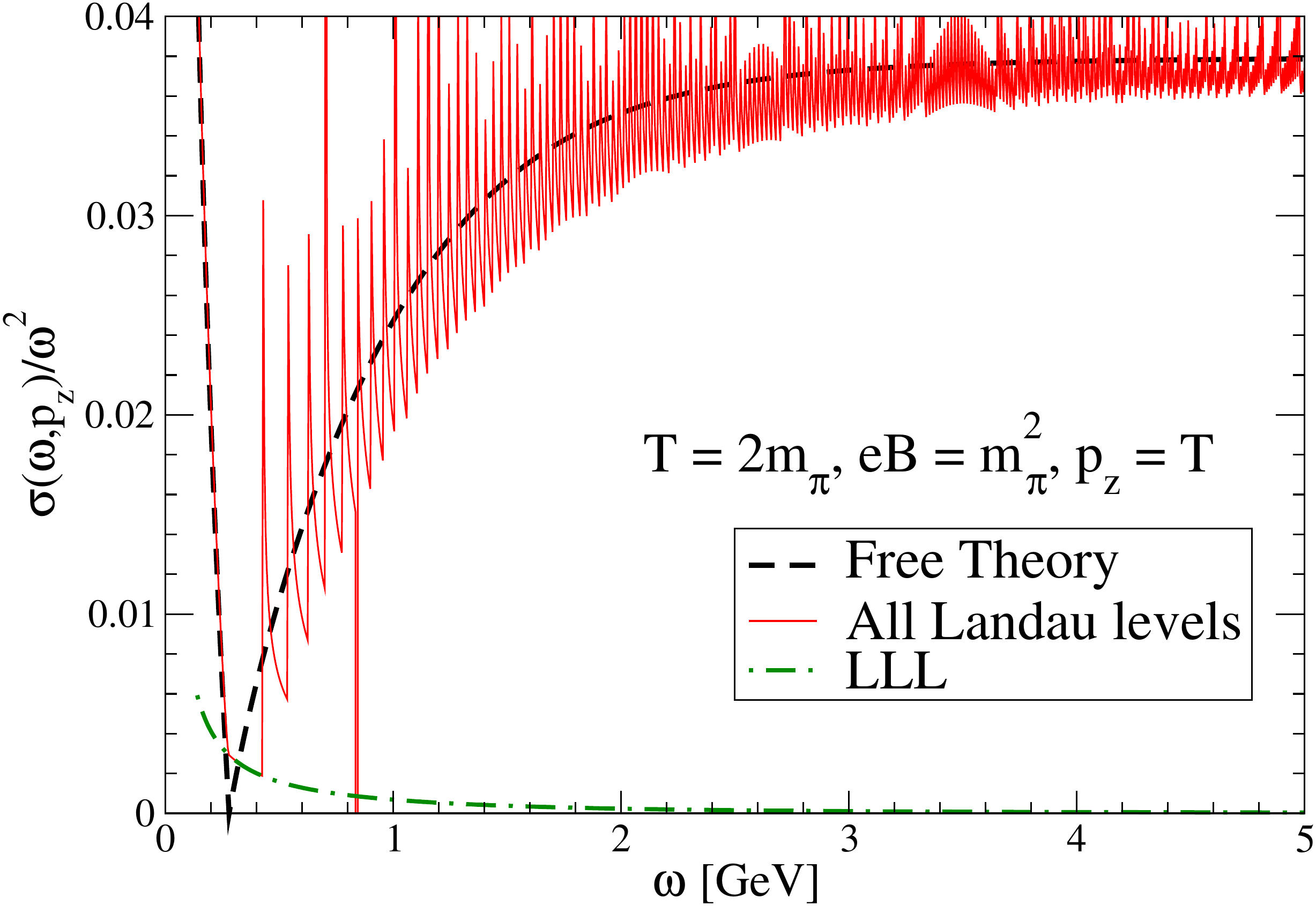}
  \end{subfigure}
\caption{\label{fig:spectpt0}Spectral function in the pseudoscalar channel
  for two different values of field strength. The meson state is assumed to be
  made up of massless $u$ quarks.} 
\end{figure}
In Fig.~\ref{fig:spectpt0}, we show spectral function in the  limit $p_\bot =
0$. The sawtooth nature of the spectral function is due to singularities at  
particle
thresholds. Physically at the point of threshold the meson is unstable with
respect decay into quark-antiquark pair. The origin of these singularities are dimensional reduction and 
infinitesimally narrow Landau levels which are artefact of lowest order of perturbation theory. 
Let us note that the peaks become narrower and the their number increases as
the intensity of the magnetic field decreases. This is easy to understand. For
a given $\omega$, the highest Landau level that contributes to the spectral
function is $r_{\rm max} = \left\lfloor(s_\parallel - 4m^2)/(8\qfb)\right\rfloor$ which increases with
dwindling magnetic field. On the other hand, in the high frequency tail of
the spectral function, the spacing between two successive peaks are given by
$\delta \omega \simeq \qfb/4\omega$ $(\omega \gg \qfb)$, which decreases when
magnetic field decreases or frequency increases.   
\begin{figure}[htbp!]
  \begin{subfigure}[b]{0.48\textwidth}
    \includegraphics[width=\textwidth]{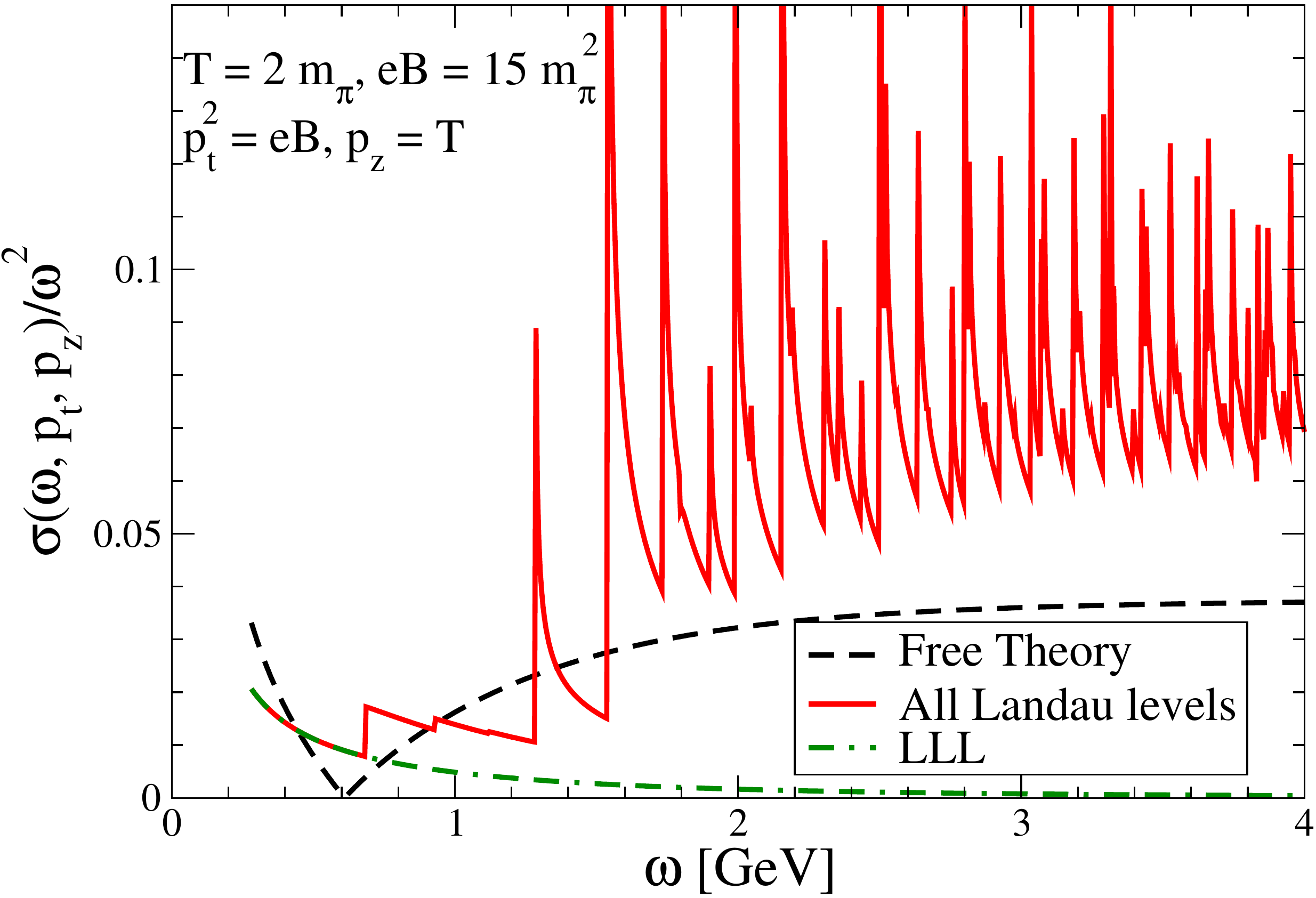}
  \end{subfigure}
  \hfill 
  \begin{subfigure}[b]{0.48\textwidth}
    \includegraphics[width=\textwidth]{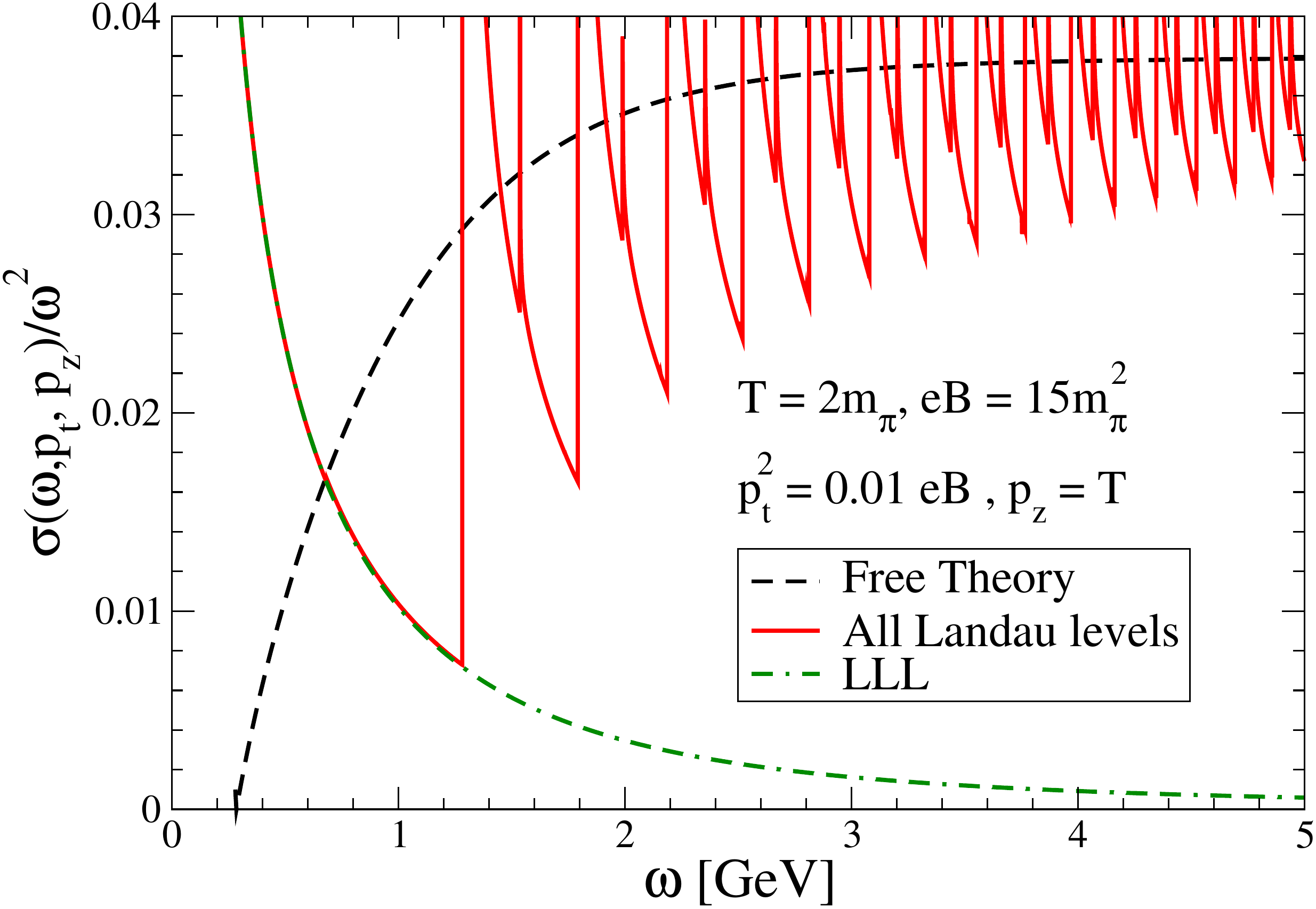}
  \end{subfigure}
\caption{\label{fig:spectptnz} Same as Fig.~\ref{fig:spectpt0} but for nonzero 
$p_\bot$ and only annihilation contribution is shown in the magnetized case. For $p_\bot \to 0$  we should get back Fig.~\ref{fig:spectpt0}. This
is shown in the Right panel.} 
\end{figure}

We show in Fig.~\ref{fig:spectptnz} the spectral function for nonzero
transverse momentum. The results are similar in nature to 
Fig.~\ref{fig:spectpt0} but more spiky and stay above than the
corresponding result at $p_\bot = 0$. These can be understood as a consequence
of availability of more decay channels at nonzero $p_\bot$. 
 
\section{Screening masses} \label{screening_mass} 

At zero momenta, the spectral function is obtained from
\begin{equation}
\label{specf_zerop} 
\sigma\left(\omega\right) = N_c \sum_{l} \left(2 - \delta_{l0}\right)
\frac{\qfb}{4\pi^2} \theta\left(\omega^2 - 8 l \qfb\right)\frac{\omega}
{\sqrt{\omega^2 - 8 l \qfb}} \tanh{\frac{\beta\omega}{4}}    
\end{equation}
We notice that far away from the threshold,  the spectral function for each 
mode consist of a frequency independent part together with power suppressed 
corrections. This is a consequence of dimensional reduction and is in 
contrast to the free field limit where spectral
function grow as $\omega^2$.  
 
Now, from \eqref{temporal_correlator} and \eqref{specf_zerop}, the temporal 
correlation function can be written as 
\begin{equation}
\label{temp_corr} 
\chi^\tau = \chi_{0}^\tau + \sum_{l > 0} \chi_l^\tau\,,
\end{equation}  
where $\chi_{0}^\tau$ is the correlator with LLL approximation. 
$\chi_{0}^\tau$ is obtained as, 
\begin{equation}
\tilde{\chi}_0^\tau = N_c \frac{\qfb}{2\pi^2 T^2} \left(1 - 2\tilde{\tau}\right) \frac{\pi}{\sin{\left(2\pi\tilde{\tau}\right)}}\,,
\end{equation} 
where $\tilde{\tau} = \tau/\beta$ and $\tilde{\chi}^\tau = \beta^3\chi^\tau$. 

\begin{figure}[!htp]
\centering{
\includegraphics[width=.5\textwidth]{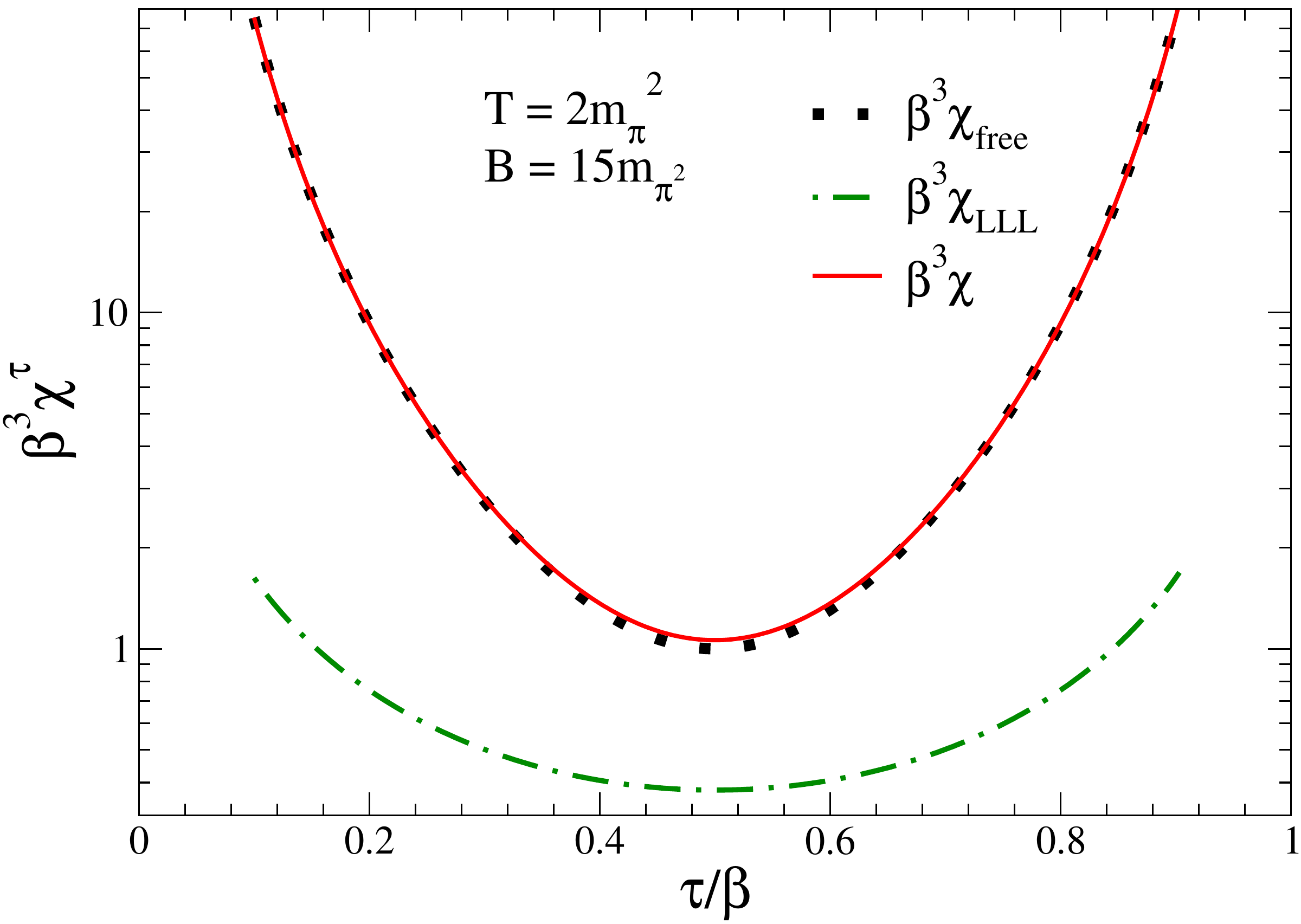}
}
\caption{\label{tcor_up} pseudoscalar correlator made up of up quark in 
  presence of a magnetic field.} 
\end{figure} 
We show the temporal correlator in the presence of a magnetic field in
Fig.~\ref{tcor_up}. The effect of the magnetic field on the spectral function is
marginal even for extreme value of the field achievable in the heavy ion
collision. Let us also note that the contribution of LLL on the temporal
correlator is small and an all order summation over the Landau levels is
necessary. The screening mass
in the time direction, on the other hand, is dictated by LLL. The screening
mass is given by $m_{\rm scr}^\tau = 2 \pi T$ which is same as in free field
theory. This behavior is understood as a consequence of $\mathcal{B}$ 
independence of LLL.

Now the immunity of $\chi^\tau$ to the exposure of $\mathcal{B}$ field can be
understood in the following way. Using Euler-Mclaurin kind of relation between the sum of series and the integral of a function~\cite{arfken_weber_2005}, it is not difficult to show that 
\begin{equation}
\label{inequality} 
\chi^\tau_{free} - \chi^\tau_{0} \le \chi^\tau \le \chi^\tau_{free} +
\chi^\tau_{0}\, 
\end{equation}  
where $\chi_{free}$ is the correlator in absence of magnetic field~\cite{Florkowski:1993bq}, 
\begin{eqnarray}
\label{free_tau_corr} 
\tilde{\chi}_{free}^\tau = \beta^3 \chi_{free}^\tau &=& \frac{N_C}{\pi^2}
\partial_{\tilde{\tau}^2} \left[\left(1 - 2\tilde{\tau}\right)
  \frac{\pi}{\sin{\left(2\pi\tilde{\tau}\right)}} \right] \nonumber \\ 
&=& 2N_c \frac{\cos{\left(2 \pi \tilde{\tau}\right)}}
{\sin^2\left(2 \pi\tilde{\tau}\right)} + N_c \pi 
\left(1 - 2 \tilde{\tau}\right) \frac{\left(1 + \cos^2\left(2 \pi\tilde{\tau}
\right)\right)}{\sin^3\left(2 \pi\tilde{\tau}\right)}\,.
\end{eqnarray} 
We have checked that inequality \eqref{inequality}  is satisfied
in our case. Thus change in the spectral function $\delta \chi = \chi  -
\chi_{free} \simeq \chi_0$ and this is small. Let us note that the derivation of \eqref{inequality} rests on the 
assumption that
the correlator can be written as a sum over  Landau levels and that it is a 
smooth
function of the quantum number of orbital motion taken as a continuous
variable. These are not overly restrictive assumptions and relations similar to
\eqref{inequality} presumably hold in general. 

The transverse plane correlator in LLL can written from
\eqref{transverse_correlator} and  \eqref{spec_lll} as
\begin{equation} 
\label{tr_corr_lll} 
\chi^{xy}\left(\vec{x}_\bot\right) = N_c \frac{\qfb}{4\pi^2} \int_0^\infty\,
\frac{d\omega}{\omega} \tan{\left(\frac{\beta\omega}{4}\right)}
\int\,\frac{d^2 p_\bot}{\left(2\pi\right)^2}\, e^{i\vec{p}_\bot\cdot\vec{x}_\bot} \,e^{-\frac{p^2_\bot}{2\qfb}}
\end{equation} 
The integration over transverse coordinates is a Gaussian one. The frequency
integration is logarithmically divergent and hence needs regularization. Explicit
from of regularization is not important at this point, it just yields a
constant, say $A$.  Now the correlator can be written as, 
\begin{equation}
\label{tr_corr_lll}
\chi^{xy}\left(\vec{x}_\bot\right) = A N_c \frac{\qfb^2}{8\pi^3}  
\exp\left({-\frac{1}{2} \qfb x^2_\bot}\right) 
\end{equation}   
The correlator decays in the transverse direction with a characteristic 
range $\sqrt{\qfb}$. But as alluded, this Gaussian falloff is not characteristics 
of typical screening behavior but rather a manifestation of magnetic
confinement.  

Using the following identity
\begin{equation}
n_f \left(x\right) = \frac{1}{2} - 2 \sum_{l = 1}^{\infty} \frac{x}{\left(2l -
  1\right)^2 \pi^2  + x^2}\,   
\end{equation} 
we obtain the longitudinal correlator  from
\eqref{transverse_correlator} and  \eqref{spec_lll} as , 
\begin{equation}
\label{lng_corr_lll} 
\chi^z\left(z\right) = N_c \frac{\qfb}{4 \pi \beta} 
\sum_{l = 1}^\infty \exp{\left(-2 \pi \frac{\left(2l - 1\right)\lvert z
    \rvert}{\beta}\right)}\,. 
\end{equation} 
At large distance $l = 1$ contribution dominates in \eqref{lng_corr_lll}. 
Thus longitudinal screening mass $m_{\rm scr}^z$ is $2\pi T$ which 
coincides with $m_{\rm scr}^\tau$.  Let us note, however,
that the asymptotic behavior of the correlator is different from free theory.  

\section{Outlook} \label{concl}
We have derived analytic expression for spectral function in  pseudoscalar (scalar)
channel in the deconfined and magnetized phase of QCD and found spatial and
temporal screening masses. While the results presented herein are nontrivial as
no approximation has been made 
regarding the kinematics or strength of the magnetic field, it would be of import to
include QCD corrections for any kind of realistic phenomenology. The point is 
that quarks interact strongly with the gluons. The strong interaction of
quarks will inevitably mix the Landau levels and fuzz the distinction between
$\parallel$ and $\perp$ spaces. 
     
From the perspective of heavy ion collision phenomenology, we belive that it
would more interesting to analyze the observables in a general background of electromagnetic
field which is inhomogeneous and time dependent. These are work 
in progress and will be reported elsewhere.

\appendix 
\section{Imaginary part of one loop self energy}\label{appendixA} 
The integration over $k_z$ can easily be done with the help of the well known
relation, 
\begin{equation}
\delta\left(f\left(x\right)\right) = \sum_{i} \frac{\delta\left(x -
  x_i\right)}{\left\lvert \frac{\partial f\left(x\right)}{\partial
    x}\right\rvert_{x_i}}\,.   
\end{equation}
$x_i$ is simple zero of $f\left(x\right)$, $f\left(x_i\right) = 0$.
The imaginary part of $\mathcal{I}_{r,s}$ is given by,
\begin{eqnarray} 
\label{eq_discont}
\Im \left(\frac{\mathcal{I}_{r,s}}{\pi}\right) &=& -\frac{1}{8\pi} 
\int \frac{dk_z}{\omega_r\left(k\right) \omega_s\left(q\right)} \biggl[\biggl(1
  - n_f\left(\omega_r\left(k\right)\right)  -
  n_f\left(\omega_s\left(q\right)\right) \biggr) \biggr\{ \delta
  \left(\omega - \omega_r\left(k\right) - \omega_s\left(q\right)\right)
   \nonumber\\ 
&-& \delta \left(\omega + \omega_r\left(k\right) +
   \omega_s\left(q\right)\right) \biggr\} + \biggl(n_f\left(\omega_r\left(k\right)\right)  - 
n_f\left(\omega_s\left(q\right)\right) \biggr) \biggl\{ \delta \left(\omega - \omega_r\left(k\right) 
 + \omega_s\left(q\right)\right)\nonumber\\ 
&-& \delta \left(\omega + \omega_r\left(k\right) 
 - \omega_s\left(q\right)\right) \biggr\} \biggr]
\end{eqnarray} 
where, $\omega_r\left(k\right) = \sqrt{k_z^2 + \epsilon_r^2}$ and 
$\omega_s\left(q\right) = \sqrt{q_z^2 + \epsilon_s^2}$. $\epsilon_r$ is
transverse mass in $r$th Landau level, $\epsilon_r^2 = m^2 + 2 r \qfb$. Let us
introduce, $\chi_{r,s} = \frac{1}{2}\left(1 + \frac{\epsilon_r^2 -
  \epsilon_s^2}{s_\parallel}\right)$, $\lambda_{rs} = \lambda_{sr} =
\frac{1}{2}\Lambda^{1/2}\left(1, \frac{\epsilon_r^2}{s_\parallel},
\frac{\epsilon_s^2}{s_\parallel}\right)$ where, $\Lambda\left(x,y,z\right) =
x^2 + y^2 + z^2 - 2xy - 2yz - 2zx$ is the triangle function. 
We also take $\omega_{rs}^\pm = \chi_{rs} \omega \pm \lambda_{rs} p_z$,
$\tilde{\omega}_{rs}^\pm = \chi_{rs} \omega \pm \lambda_{rs} \lvert
p_z\rvert$. The imaginary part can be written as,  
\begin{eqnarray}
\Im \left(\frac{\mathcal{I}_{r,s}}{\pi}\right) &=& - \theta\left(s_\parallel -
\left(\epsilon_r + \epsilon_s\right)^2\right) \frac{1}{8\pi s_\parallel
  \lambda_{rs}} \biggl(2 - n_f \left(\omega_{rs}^+\right)  - n_f
\left(\omega_{rs}^-\right) - n_f \left(\omega_{sr}^+\right)  - n_f
\left(\omega_{sr}^-\right)\biggr) \nonumber \\ 
&+& \biggl( \theta
\left(s_\parallel\right) - \theta\left(s_\parallel -\left(\epsilon_r -
\epsilon_s\right)^2\right)\biggr) \frac{1}{4\pi s_\parallel
  \lambda_{rs}} \biggl(2 - n_f \left(\omega_{rs}^+\right)  - n_f
\left(\omega_{rs}^-\right) - n_f \left(\omega_{sr}^+\right)  - n_f
\left(\omega_{sr}^-\right)\biggr) \nonumber \\ 
 &-& \theta^\star\left(\epsilon_r - \epsilon_s\right) \theta \left(-s_\parallel\right) 
\frac{1}{8\pi s_\parallel \lambda_{rs}} \biggl(n_f \left(\tilde{\omega}_{rs}^+\right)
- n_f \left(-\tilde{\omega}_{sr}^-\right)\biggr) \nonumber \\  
&-& \theta^\star\left(\epsilon_s - \epsilon_r\right) \theta \left(-s_\parallel\right) 
\frac{1}{8\pi s_\parallel \lambda_{rs}} \biggl(n_f \left(\tilde{\omega}_{sr}^+\right)
- n_f \left(-\tilde{\omega}_{rs}^-\right)\biggr) \nonumber \\ 
&-& \delta_{\epsilon_r,\epsilon_s}\,\theta
\left(-s_\parallel\right)  \frac{1}{4\pi s_\parallel \lambda_{rs}} \biggl(n_f \left(\tilde{\omega}_{rr}^+\right)
- n_f \left(-\tilde{\omega}_{rr}^-\right)\biggr)\,,
\end{eqnarray} 
where $\theta*\left(x\right)$ is Heaviside theta function with the condition
that $\theta^*\left(0\right) = 0$. 
 
\section{Evaluation of \ensuremath{\mathcal{T}} intergrals}
The $\mathcal{T}$ integrals in the main text have following structure,    
\begin{equation}
\label{intT}
\mathcal{T}_{m,n}^{\alpha,\beta, \gamma} = \left(-1\right)^{m+n} \int
\frac{d^2k_\perp}{\left(2\pi\right)^2} e^{-\left(\rho_k + \rho_q\right)}
L_m^\alpha\left(2 \rho_k\right) L_n^\beta\left(2 \rho_q\right)  
\left(\vec{k}\cdot\vec{q}\right)^\gamma\,.
\end{equation} 
Let us scale momentum variables as $x = \frac{2 k_\perp^2}{|q_f B|}$ and $\xi = \frac{2
  p_\perp^2}{|q_f B|}$ in \eqref{intT}. We can write 
\begin{equation}
\mathcal{T}_{m,n}^{\alpha,\beta, \gamma} = \left(-1\right)^{m+n} \frac{\lvert q_f \mathcal{B}\rvert}{16
  \pi^2} \left(\frac{\lvert q_f \mathcal{B}\rvert}{2}\right)^\gamma 
e^{-\frac{\xi}{2}} 
\mathcal{\chi}^{\alpha, \beta, \gamma}_{m,n}\,, 
\end{equation} 
where 
\begin{equation}
\label{chi_def}
\mathcal{\chi}^{\alpha, \beta, \gamma}_{m,n} = \int_0^\infty dx \int_0^{2\pi} d\phi\, 
e^{-\left(x - \sqrt{x\xi} \cos{\phi}\right)} L_m^\alpha\left(x\right)  L_n^\beta\left(x + \xi - 2\sqrt{x\xi}
\cos{\phi}\right)
\left(x - \sqrt{x\xi} \cos{\phi}\right)^\gamma\,.
\end{equation} 
\eqref{chi_def} is defined for arbitrary positive values of $\alpha$, $\beta$ or 
$\gamma$.  For our purpose, we need to evaluate a small
subset of this where $\alpha = \beta = \gamma \in \left\{0,1\right\}$.  
When $\xi = 0$, $\mathcal{\chi}_{m,n}^\alpha = \mathcal{\chi}_{m,n}^{\alpha,\alpha,\alpha}$  
express the orthogonality relation for Laguerre polynomials,    
\begin{equation}
\label{normalization_integral}
\mathcal{\chi}^{\alpha, \alpha, \alpha}_{m,n} = 2 \pi \frac{\Gamma\left(\alpha + n +
  1\right)}{\Gamma\left(n + 1\right)} \delta_{m,n}\,.  
\end{equation}  
For nonzero value of $\xi$ deterministic numerical integrators perform fairly
well to evaluate diagonal elements $\mathcal{\chi}_{mm}^\alpha$ for moderate value of
$\xi$.  For extreme values of $\xi$ or when $\lvert m - n \rvert$ is large, success
of cubature routines are uncertain due to oscillatory nature of the
integrand in \eqref{chi_def}. We can, however, circumvent this problem by expanding $\mathcal{\chi}$
in a polynomial basis. Since $\mathcal{\chi}$ is an analytic function of $\xi$, such an
expansion is always possible and it provides a fast, stable and accurate 
method for the numerical evaluation of \eqref{chi_def}. In addition to this,
analytic expressions for $\mathcal{\chi}$ may prove to  be useful to test the
accuracy of the results obtained from approximate forms of propagators.

We assume a Laguerre-Fourier expansion of $\mathcal{\chi}$   
\begin{equation}
\mathcal{\chi} \left(\xi\right) = \sum_{l = 0}^\infty c_l^\delta L_l^\delta \left(\xi\right)\,,
\end{equation} 
and our task boils down to finding out the coefficients
$c_l^\delta$ which we will do in a heuristic way. 

Let us first consider $\mathcal{\chi}_{mn}^0$. The strategy here is to exponentiate the angular dependence 
in one of the Laguerre polynomial in \eqref{chi_def} using a nice addition 
formula due to Bateman~\cite{bateman1932partial},
\begin{equation}
\label{bateman_expansion}
e^{\sqrt{x\xi} e^{i\phi}} L_n^\beta\left(x + \xi - 2\sqrt{x\xi}\cos{\phi}\right) = 
\sum_{k = 0}^\infty \left(\sqrt{x\xi} e^{i \phi}\right)^{k-n}  \frac{n!}{k!}
L_n^{k-n} \left(x\right) 
L_n^{k-n}\left(\xi\right)\,. 
\end{equation}  
We multiply both sides of \eqref{bateman_expansion} by $e^{-i\sqrt{x\xi}
  \cos{\phi}}$ and substitute it in \eqref{chi_def}. The innermost angular 
integration can be performed using Sommerfeld's representation of Bessel 
function, 
\begin{equation}
J_\nu \left(u\right) = \frac{1}{2\pi} \int_0^{2\pi} d\phi\, e^{i u
  \sin{\phi} - i\nu \phi}\,, \quad  \text{$\nu$ is an integer}.
\end{equation} 
The $x$ integration then can be done with the following identity
\cite{kolbig1996hankel}, 
\begin{eqnarray}
&&\int_0^\infty\, dx\, e^{-x} \left(\sqrt{x}\right)^{\gamma + \lambda} J_{\gamma +
  \lambda} \left(b \sqrt{x}\right) L_m^\gamma\left(x\right)  
L_n^\lambda\left(x\right)\nonumber \\
&&\,\,\, = \left(-1\right)^{m+n}  \left(\frac{b}{2}\right)^{\gamma + \lambda}
 e^{-\frac{b^2}{4}}\, L_m^{\gamma + m -n}\left(\frac{b^2}{4}\right) 
L_n^{\lambda + n - m}\left(\frac{b^2}{4}\right)\,.
\label{erdelyi_hankel}
\end{eqnarray} 
So we have, 
\begin{eqnarray}
\mathcal{\chi}^0_{m,n} &=& \left(-1\right)^{m+n} 2\pi \Gamma\left(n + 1\right)
\left(\frac{\xi}{2}\right)^{-n} e^{-\frac{\xi}{4}} L_n^{m - n}\left(\frac{\xi}{4}\right) \nonumber \\
&\times& \sum_{k=0}^{n} \frac{1}{k!} \left(\frac{\xi}{2}\right)^{k}
L_n^{k-n}\left(\xi\right) L_m^{m - n}\left(\frac{\xi}{4}\right)\,.
\label{chi01}
\end{eqnarray}
We can further simplify \eqref{chi01} by using the following 
identity~\cite{erdelyi1940transformation}, 
\begin{equation}
\sum_{k = 0}^{\infty} L_m^{\left(k + \alpha\right)} \left(x\right)
L_n^{\left(k + \beta\right)} \left(y\right)
\frac{z^k}{k!} = e^z \sum_{k = 0}^{\text{min}(m,n)} L_{m - k}^{\left(k +
  \alpha\right)} \left(x - z\right)
L_{n - k}^{\left(k + \beta\right)} \left(y - z\right)
\frac{z^k}{k!}\,, 
\end{equation} 
together with the fact that $L_\nu^{-\nu} \left(x\right) = (-x)^\nu/\nu!$. The
upshot is an amazingly simple expression for $\mathcal{\chi}^0_{m,n}$, 
\begin{equation}
\label{chi0int}
\mathcal{\chi}^0_{m,n} = \left(-1\right)^m \frac{2 \pi e^{\frac{\xi}{4}}}{\Gamma\left(m
  + 1\right)} U\left(-m, 
- m + n + 1, \frac{\xi}{4}\right) L_n^{m - n}\left(\frac{\xi}{4}\right)\,, 
\end{equation} 
where $U(a, b, z)$ is Tricomi's confluent hypergeometric function. 

The expression for $\mathcal{\chi}_{m,n}^1$ is somewhat complicated. To evaluate it, we
start with the following generalization of \eqref{bateman_expansion}\cite{koornwinder1977addition}, 
\begin{equation}
\label{koornwinder_expansion}
e^{\sqrt{x\xi} e^{i\phi}} L_n^\beta\left(x + \xi - 2\sqrt{x\xi}\cos{\phi}\right) = 
\sum_{k = 0}^\infty \sum_{s =0}^n \mathcal{C}^\beta_{nks} \left(\sqrt{x\xi}\right)^{k +
  s}  L_{n - s}^{\beta + k + s} \left(x\right) L_{n - s}^{\beta + k +
  s}\left(\xi\right) e^{i\phi\left(k - s\right)}\,, 
\end{equation} 
where
\begin{equation}
\mathcal{C}^\beta_{nks} = \frac{\left(\beta + k + s\right) \Gamma\left(\beta +
  s\right) \Gamma\left(\beta + k\right) \Gamma\left(n - s + 1\right)}
 {\Gamma\left(\beta\right) \Gamma\left(s + 1\right) \Gamma\left(k + 1\right) 
\Gamma\left(\beta + n + k + 1\right)}\,, \beta \neq 0 . 
\end{equation}
As in the earlier case, \eqref{koornwinder_expansion} allows to perform the
angular integration. The $x$ integration is likewise 
recast in the form of a Hankel transform,  a convenient closed form expression
for which can be derived as,   
\begin{eqnarray}
\mathcal{H}_{m,n}^{\mu,\nu,\alpha,\beta} &=& \int_0^\infty\, dx\, e^{-x}
x^{\mu + \frac{\nu}{2}}  L_m^\alpha\left(x\right)  
L_n^\beta\left(x\right) J_\nu \left(\sqrt{x \xi}\right)\nonumber \\
&=& \left(-1\right)^{m+n}  \Gamma\left(\mu + 1\right) 
\left(\frac{\sqrt{\xi}}{2}\right)^\nu 
 e^{-\frac{\xi}{4}} \sum_{r = 0}^\mu \sum_{l = 0}^\gamma
 \frac{\left(-1\right)^r}{r!} \binom{\mu + \nu}{\mu  - r} \binom{\gamma}{l}
 \nonumber \\ 
&\times& L_n^{\alpha + m + l - n}\left(\frac{\xi}{4}\right) 
L_{m + l -\gamma}^{\beta + n + \gamma - l - m}\left(\frac{\xi}{4}\right)\,, 
\label{hankel_pc} 
\end{eqnarray} 
where $\gamma = \mu + \nu -\alpha -\beta + r$ and $\mu + \nu > -1$. For $\gamma < 0$, the uppper
limit of $l$ summation in \eqref{hankel_pc} has to be replaced by
$\infty$. For $\mu = 0, \nu = \alpha + \beta$ \eqref{hankel_pc} reduces to
\eqref{erdelyi_hankel}. 

After a short algenbra $\mathcal{\chi}^1_{m,n}$ can be expressed as,   
\begin{eqnarray}
\mathcal{\chi}^1_{m,n} &=&  {^a\mathcal{\chi}}^1_{m,n} - {^b\mathcal{\chi}}^1_{m,n}\,,\nonumber \\ 
{^a\mathcal{\chi}}^1_{m,n} & = & 2 \pi \sum_{k = 0}^\infty \sum_{s = 0}^n
\mathcal{C}^\beta_{nks} \left(\sqrt{\xi}^{k + s}\right) \mathcal{H}_{m,n-s}^{s
  + 1,k - s,1, k + s + 1}\,,\nonumber \\
{^b\mathcal{\chi}}^1_{m,n} & = & 2 \pi \sum_{k = 0}^\infty \sum_{s = 0}^n
\mathcal{C}^\beta_{nks} \left(\sqrt{\xi}^{k + s}\right) \left(k - s\right) \mathcal{H}_{m,n-s}^{s,k - s,1, k + s + 1}\,.
\label{chi1} 
\end{eqnarray}  

 

\bibliographystyle{JHEP}
\bibliography{corr_fn_refs}

\end{document}